%% file: main.tex
\documentclass{article}

\usepackage{latexsym}
\usepackage{graphics}

\input{prelude}
\input{abbreviations}

\newenvironment{alist}[1]
  {\begin{list}{#1}{\parsep=0ex\topsep=0ex\itemsep=0ex}}
  {\end{list}}

\newenvironment{blist}[1]
  {\begin{list}{#1}{\parsep=0ex\topsep=0ex\itemsep=0ex
                    \leftmargin=5mm\itemindent=-5mm}}
  {\end{list}}

\newcommand{\term}{Term}
\newcommand{\neutr}{N\!eutr}

\newcommand{\var}{V\!ar}
\newcommand{\varstar}{\var^\star}
\newcommand{\varbox}{\var^\Box}

\newcommand{\pos}{Pos}
\newcommand{\fv}{FV}
\newcommand{\bv}{BV}
\newcommand{\sn}{SN}

\newcommand{\dom}{dom}

\newcommand{\type}{Type}

\newcommand{\kind}{Kind}

\newcommand{\const}{{\cal C}}
\newcommand{\ind}{Ind}

\newcommand{\inter}[1]{\interxx{}{#1}}

\newcommand{\superterm}{\,\rhd\,}
\newcommand{\supertermgams}[2]{\,\rhd\!\!_{_{#1}\!,{#2}}\,}

\newcommand{\supertermeq}{\,\unrhd\,}
\newcommand{\supertermeqgams}[2]{\,\unrhd_{_{#1}\!,{#2}}\,}

\newcommand{\interxx}[2]{[\![#2]\!]_{#1}}
\newcommand{\crit}[2]{\chi\!_{_{#1}}^{#2}}
\newcommand{\critint}[1]{\phi_{\!f,_\Gamma}^{#1}}
\newcommand{\rcint}[2]{\Phi_{\!f\!,_{#1}}^{#2}}

\newcommand{\imply}{\,\Rightarrow\,}

\newcommand{\arrow}{\lrps{}{}}

\newcommand{\tarrow}{\!\rightarrow\!}

\newcommand{\arrowstar}{\arrow\!\!\!^*}

\newcommand{\these}{\,\vdash\,}

\newcommand{\ded}[3]{#1\!\these\!#2\!:\!#3}

\newcommand{\decl}[2]{#1\!\!:\!\!#2}
\newcommand{\declb}[3]{#1,\decl{#2}{#3}}
\newcommand{\dedb}[5]{\declb{#1}{#2}{#3}\!\these\!#4\!:\!#5}

\newcommand{\cfrac}[2]{\frac{\mbox{$#1$}}{\mbox{$#2$}}}
\newcommand{\barre}{\;|\;}

\newcommand{\starbox}{\{\star,\Box\}}

\newcommand{\gams}{\Gamma\!,\!s}
\newcommand{\gamR}{\Gamma\!_R}
\newcommand{\asort}[1]{{\tt #1}}

\newcommand{\bool}{{\tt bool}}
\newcommand{\true}{{\tt true}}
\newcommand{\false}{{\tt false}}
\newcommand{\nat}{{\tt nat}}
\newcommand{\onat}{{\tt 0}} 
\newcommand{\snat}{{\tt s}} 

\newcommand{\sortlist}[1]{{\tt list}_{#1}}
\newcommand{\nil}[1]{{\tt nil}_{#1}}
\newcommand{\cons}[1]{{\tt cons}_{#1}}

\newcommand{\ord}{{\tt ord}}
\newcommand{\oord}{{\tt 0}_\ord}
\newcommand{\sord}{{\tt s}_\ord}
\newcommand{\limord}{{\tt lim}_\ord}

\newcommand{\ifthenelse}[1]{{{\tt if}_{#1}}}
\newcommand{\plusnat}{{\tt +}} 

\newcommand{\append}[1]{{\tt append}_{#1}}
\newcommand{\map}[1]{{\tt map}_{#1}}
\newcommand{\ack}{{\tt ack}}

\newcommand{\boolrec}{{\tt rec}_\bool^t}
\newcommand{\natrec}{{\tt rec}_\nat^t}

\newcommand{\ordrec}{{\tt rec}_\ord^t}

\newcommand{\sig}[2]{{{\cal F}_{#1}^{#2}}}

\newcommand{\sort}{\Sort}
\newcommand{\algtype}{{\cal T}_{\sort}}

\newcommand{\pro}[3]{\Pi #1\!\!:\!\!#2.#3}
\newcommand{\abs}[3]{\lambda #1\!\!:\!\!#2.#3}

\newcommand{\app}[2]{#1~#2}

\newcommand{\remp}[3]{{#1[#2]_{#3}}}
\newcommand{\subt}[2]{{#1}|_{#2}}

\newcommand{\lexx}[1]{\,{#1}_{lex}\,}

\newcommand{\sub}[2]{\{{#1}\!\mapsto\!{#2}\}}
\newcommand{\subvec}[1]{\{\vec{x}\!\mapsto\!\vec{#1}\}}

\newcommand{\Y}[1]{\cal Y}

\newcommand{\CS}[2]{{\cal CC}_{#1}(#2)}
\newcommand{\Sort}{{\cal S}}

\newcommand{\gtF}{>_{\cal F}}

\newcommand{\ltF}{<_{\cal F}}

\begin{document}

\input{title}

\input{introduction}

\input{language}

\input{strong-normalization}

\input{conclusion}

\vspace{2mm}
\noindent {\bf Acknowledgements:} We want to thank Maribel
Fern\'andez for her careful reading, and the useful remarks by the
anonymous referees.

\vspace{-3mm}

\input{biblio}
\end{document}

%% file: prelude.tex
\newtheorem{definition}{Definition}

\newtheorem{theorem}[definition]{Theorem}

\newtheorem{proposition}[definition]{Proposition}

\newtheorem{corollary}[definition]{Corollary}

\newtheorem{lemma}[definition]{Lemma}

\newtheorem{example}[definition]{Example}

\newenvironment{proof}{Proof:}{}
\newcommand{\qed}{\cqfd}

\newcommand{\FDEF}{\end{definition}}
\newcommand{\PROP}{\begin{proposition}}
\newcommand{\FPROP}{\end{proposition}}
\newcommand{\COR}{\begin{corollary}}
\newcommand{\FCOR}{\end{corollary}}
\newcommand{\EX}{\begin{example}}
\newcommand{\FEX}{\end{example}}
\newcommand{\FTH}{\end{theorem}}
\newcommand{\LEM}{\begin{lemma}}
\newcommand{\FLEM}{\end{lemma}}

%% file: abbreviations.tex
\newcommand{\comment}[1]{}

\newcommand{\cqfd}{\hfill $\Box$}

\newcommand{\turnstyle}{~\vdash~}

\newcommand{\ra}{\rightarrow}

\newcommand{\Nat}{\mbox{$\mbox{I}\!\mbox{N}$}}

\newcommand{\Pos}[1]{{\cal P}os({#1})}

\newcommand{\eqps}[2]{\mathop{\longleftrightarrow}^{#1}_{#2}}

\newcommand{\lrps}[2]{\mathop{\longrightarrow}^{#1}_{#2}}

\newcommand{\gesubt}{\unrhd}

\newcommand{\mul}[1]{\mathop{#1}_{mul}}

\newenvironment{ruleset}
               {$$ \begin{array}{llcr}}{\end{array} $$ }

\newcommand{\Farc}[2]{\displaystyle \frac{#1}{#2}}



%% file: title.tex
\title{{\bf The Calculus of Algebraic Constructions}\thanks{This work
was partly supported by the Grants-in-aid for Scientific Research of
Ministry of Education, Science and Culture of Japan, and the
Oogata-kenkyuu-jyosei grant of Keio University.}}

\vspace{25mm}

\author{Fr\'ed\'eric Blanqui{$^{\dag}$}, Jean-Pierre Jouannaud{$^{\dag}$}
and Mitsuhiro Okada{$^{\ddag}$}\\
\small $^\dag$~LRI, CNRS UMR 8623 et Universit\'e Paris-Sud\\
\small B\^at. 405, 91405 Orsay Cedex, France\\
\small $^\ddag$~Department of Philosophy, Keio University,\\
\small 108 Minatoku, Tokyo, Japan\\
\small Tel: +33-1-69156905 ~~~~ FAX: +33-1-69156586 ~~~~
Tel-FAX:+33-1-43212975}

\date{}

\maketitle

\vspace{-3mm}
\noindent {\bf Abstract :} This paper is concerned with the
foundations of the Calculus of Algebraic Constructions (CAC), an
extension of the Calculus of Constructions by inductive data types.
CAC generalizes inductive types equipped with higher-order primitive
recursion, by providing definitions of functions by pattern-matching
which capture recursor definitions for arbitrary non-dependent and
non-polymorphic inductive types satisfying a strictly positivity
condition. CAC also generalizes the first-order framework of abstract
data types by providing dependent types and higher-order rewrite
rules.

%% file: introduction.tex
\section{Introduction}

Proof assistants allow one to build complex proofs by using macros,
called tactics, which generate proof terms representing the sequence
of deduction rules used in the proof. These proof terms are then
``type-checked'' in order to ensure the correct use of each deduction
step. As a consequence, the correctness of the proof assistant, hence
of the verification itself, relies solely on the correctness of the
type-checker, but not on the tactics themselves. This approach has a
major problem: proof objects may become very large. For example,
proving that $0+100$ equals its normal form $100$ in some encoding of
Peano arithmetic will generate a proof of a hundred steps, assuming
$+$ is defined by induction on its second argument. Such proofs occur
in terms, as well as in subterms of a dependent type. Our long term
goal is to cure this situation by restoring the balance between
computations and deductions, as argued in~\cite{dowek98inria3400}. The
work presented in this paper intends to be a first important step
towards this goal. To this end, we will avoid encodings by
incorporating to the Calculus of Constructions (CC)~\cite{coquand88ic}
user-defined function symbols defined by sets of first and
higher-order rewrite rules. These rules will be used in conjunction
with the usual proof reduction rule that reduces subterms in dependent
types:
\begin{center}
$\Farc{\Gamma\turnstyle M:T\quad T\eqps{*}{R\cup\beta}T'}
{\Gamma\turnstyle M:T'}$
\end{center}
Since the pioneer work by Breazu-Tannen in 1988~\cite{breazu88lics} on
the confluence of the combination of the simply-typed
$\lambda$-calculus with first-order algebraic rewriting, soon
followed, as for the strong normalization, by Breazu-Tannen and
Gallier~\cite{breazu91tcs} and, independently, by
Okada~\cite{okada89sac}, this question has been very active. We
started our program at the beginning of the decade, by developing the
notion of abstract data type system~\cite{jouannaud97tcs}, in which
the user defined computations could be described by using rewrite
rules belonging to the so-called {\em General Schema}, a
generalization of higher-order primitive recursion. This work was done
in the context of a bounded polymorphic type discipline, and was later
extended to CC~\cite{barbanera97jfp}.

In~\cite{blanqui98tcssub}, we introduced, in the context of the
simply-typed $\lambda$-calculus, a new and more flexible definition of
the General Schema to capture the rewrite rules defining recursors for
strictly positive inductive types~\cite{coquand88colog}, problem left
open in~\cite{jouannaud97tcs}. In this paper, we similarly equip CC
with non-dependent and non-polymorphic inductive types, and first and
higher-order rewriting. Our main result is that this extension is
compatible with CC.

In~\cite{coquand88colog}, strictly positive inductive types can be
dependent and polymorphic. Hence, further work will be needed to reach
the expressive power of the Calculus of Inductive
Constructions~\cite{werner94these}, implemented in the Coq proof
assistant~\cite{coq98}, all the more so since it handles strong
elimination, that is the possibility to define types by induction.
But our new General Schema seems powerful and flexible enough to be
further extended to such a calculus, hence resulting in to a simpler
strong normalization proof.

As a consequence of our result, it will become possible to develop a
new version of the Coq proof assistant, in which the user may define
functions by pattern-matching and then develop libraries of decision
procedures using this kind of functional style. Ensuring the
consistency of the underlying proof theory requires a proof that the
user-defined rules obey the General Schema, a task that can be easily
automated. Note also that, since most of the time, when one develops
proofs, the efficiency of rewriting does not really matter, the
type-checker of the proof development system can be kept small and not
too difficult to certify, hence conforming to the idea of relying on a
small easy-to-check kernel.


%% file: language.tex

\section{Definition of the calculus}


\subsection{Syntax}


\begin{definition}[Algebraic types]
Given a set $\sort$ of {\em sorts}, we define the sets $\algtype$ of
{\em algebraic types}:
\vspace{-4mm}
\begin{center}
$s:= \asort{s}\barre (s\tarrow s)$
\end{center}
\vspace{-2mm}
where $\asort{s}$ ranges over $\sort$ and
$\tarrow$ associates to the right such that $s_1\tarrow(s_2\tarrow s_3)$
can be written $s_1\tarrow s_2\tarrow s_3$.
An algebraic type $s_1\tarrow\ldots\tarrow s_n$ is {\em first-order}
if each $s_i$ is a sort, otherwise it is {\em higher-order}.
\end{definition}


\begin{definition}[Constructors]
  We assume that each sort $\asort{s}$ has an associated set
  $\const(\asort{s})$ of {\em constructors}. Each constructor $C$ is
  equipped with an algebraic type $\tau(C)$ of the form
  $s_1\tarrow\ldots\tarrow s_n\tarrow \asort{s}$; $n$ is called the
  {\em arity} of $C$, and $\asort{s}$ its {\em output type}. We denote
  by $\const^n$ the set of constructors of arity $n$.
  
  A constructor $C$ is {\em first-order} if its type is first-order,
  otherwise it is {\em higher-order}. Constructor declarations define
  a quasi-ordering on sorts: $\asort{s}\ge_\sort \asort{t}$ if and
  only if $\asort{t}$ occurs in the type of a constructor belonging to
  $\const(\asort{s})$.  In the following, we will assume that
  $>_\sort$ is well-founded, ruling out mutually inductive sorts.
\end{definition}


\begin{definition}[Algebraic signature]
  Given a non empty sequence $s_1,\ldots,s_n,s$ of algebraic types, we
  denote by $\sig{s_1,\ldots,s_n,s}{}$ the set of {\em function
    symbols} of {\em arity} $n$, of {\em type}
  $\tau(f)=s_1\tarrow\ldots\tarrow s_n\tarrow s$ and of {\em output
    type} $s$.  We will denote by $\sig{}{n}$ the set of function
  symbols of arity $n$, and by $\sig{}{}$ the set of all function
  symbols.  Function symbols with a first-order (resp. higher-order)
  type are called {\em first-order} (resp. {\em higher-order}).
\end{definition}


Here are familiar examples of sorts and functions:

(i) the sort $\bool$ of booleans whose constructors are $\true:\bool$
and $\false:\bool$; $\ifthenelse{t}$ of arity 3 is a defined function
of type \bool\ $\tarrow$ $t$ $\tarrow$ $t$ $\tarrow$ $t$, for any
algebraic type $t$;

(ii) the sort $\nat$ of natural numbers whose constructors are
$\onat:\nat$ and $\snat:\nat \tarrow \nat$; $\plusnat$ of arity 2 is a
defined function of type \nat\ $\tarrow$ \nat\ $\tarrow$ \nat;

(iii) the sort $\sortlist{t}$ of lists of elements of an algebraic
type $t$ whose constructors are $\nil{t}:\sortlist{t}$ and $\cons{t}:
t \tarrow \sortlist{t} \tarrow \sortlist{t}$; $\append{t}$ of arity 2
is a defined function of type $\sortlist{t}$ $\tarrow$ $\sortlist{t}$
$\tarrow$ $\sortlist{t}$, while $\map{t,t'}$ of arity 2 is a defined
function of type ($t$ $\tarrow$ $t'$) $\tarrow$ $\sortlist{t}$
$\tarrow$ $\sortlist{t'}$;

(iv) the sort $\ord$ of ordinals whose constructors are $\oord:\ord$,
$\sord:\ord \tarrow \ord$ and $\limord:(\nat \tarrow \ord) \tarrow
\ord$.


\begin{definition}[Terms]
The set $\term$ of CAC {\em terms} is inductively defined as:
\[a := x\barre \asort{s}\barre \star\barre \Box\barre
\abs{x}{a}{a}\barre \pro{x}{a}{a}\barre (\app{a}{a})\barre
C(a_1,\ldots,a_n)\barre f(a_1,\ldots,a_n) \]
where $\asort{s}$ ranges over $\sort$, $C$ over $\const^n$,
$f$ over $\sig{}{n}$ and $x$ over $\var$, a set of variables made of
two disjoint infinite sets $\varbox$ and $\varstar$.
The application ($\app{a}{b}$) associates to the left such that
$\app{(\app{a_1}{a_2})}{a_3}$ can be written
$\app{\app{a_1}{a_2}}{a_3}$.
The sequence of terms $a_1\ldots a_n$ is
denoted by the vector $\vec{a}$ of length $|\vec{a}|=n$.
A term $C(\vec{a})$ (resp. $f(\vec{a})$) is said to be
{\em constructor headed} (resp. {\em function headed}).
\end{definition}

After Dewey, the set $\pos(a)$ of {\em positions} in a term $a$ is a
language over the alphabet $\Nat^+$ of strictly positive natural
numbers. Note that abstraction and product have two arguments, the
type and the body. The {\em subterm} of a term $a$ at position
$p\in\pos(a)$ is denoted by $\subt{a}{p}$ and the term obtained by
replacing $a|_p$ by a term $b$ is written $\remp{a}{b}{p}$. We write
$a\gesubt b$ if $b$ is a subterm of $a$.


We note by $\fv(a)$ and $\bv(a)$ the sets of respectively free and
bound variables occurring in a term $a$, and by $\var(a)$ their union.
By convention, {\em bound and free variables will always be assumed
  different}.  As in the untyped $\lambda$-calculus, terms that only
differ from each other in their bound variables will be identified, an
operation called {\em $\alpha$-conversion}.  A {\em substitution}
$\theta$ of domain $\dom(\theta)= \{\vec{x}\}$ is written $\{\vec{x}
\mapsto \vec{b}\}$. Substitutions are written in postfix notation, as
in $a\theta$.

Finally, we traditionally consider that $(b~\vec{a})$,
$\abs{\vec{x}}{\vec{a}}{b}$ and $\pro{\vec{x}}{\vec{a}}{b}$, denote
all three the term $b$ if $\vec{a}$ is the empty sequence, and the
respective terms $(\ldots((b~a_1)~a_2)$ $\ldots a_n)$,
$\abs{x_1}{a_1}{( \abs{x_2}{a_2}{( \ldots( \abs{x_n}{a_n}{b})
    \ldots)})}$ and $\pro{x_1}{a_1}{( \pro{x_2}{a_2}{( \ldots(
    \pro{x_n}{a_n}{b}) \ldots)})}$ otherwise.  We also write $a\tarrow
b$ for the term $\pro{x}{a}{b}$ when $x\not\in\fv(b)$. This
abbreviation allows us to see algebraic types as terms of our
calculus.


\subsection{Typing rules}


\begin{figure}[t]
\caption{Typing rules of CAC\label{fig-typ-rules}}
\vspace{-2mm}
\begin{tabular}{rcl}
\\ (ax) & $\ded{}{\star}{\Box}$ \\

\\ (sort) & $\ded{}{\asort{s}}{\star}$ & $(\asort{s}\in\sort)$ \\

\\ (var) & $\cfrac{\ded{\Gamma}{c}{p}}{\ded{\Gamma, x:c}{x}{c}}$
  & $(x\in\var^p\setminus\dom(\Gamma),\, p\in\starbox)$ \\

\\ (weak) & $\cfrac{\ded{\Gamma}{a}{b} \quad
  \ded{\Gamma}{c}{p}}{\ded{\Gamma, x:c}{a}{b}}$
  & $(x\in\var^p\setminus\dom(\Gamma),\, p\in\starbox)$ \\

\\ (cons) & $\cfrac{\ded{\Gamma}{a_1}{s_1}\quad\ldots\quad
  \ded{\Gamma}{a_n}{s_n}}{\ded{\Gamma}{C(a_1,\ldots,a_n)}{\asort{s}}}$
  & $(C\in\const^n,\,\tau(C)=s_1\tarrow\ldots\tarrow
  s_n\tarrow \asort{s})$\\

\\ (fun) & $\cfrac{\ded{\Gamma}{a_1}{s_1}\quad\ldots\quad
  \ded{\Gamma}{a_n}{s_n}}{\ded{\Gamma}{f(a_1,\ldots,a_n)}{s}}$
  & $(f\in\sig{s_1,\ldots,s_n,s}{},\,n\ge 0)$ \\

\\ (abs) & $\cfrac{\ded{\Gamma, x:a}{b}{c}
  \quad \ded{\Gamma}{(\pro{x}{a}{c})}{q}}
  {\ded{\Gamma}{(\abs{x}{a}{b})}{(\pro{x}{a}{c})}}$
  & $(x\not\in\dom(\Gamma),\,q\in\starbox)$ \\

\\ (app) & $\cfrac{\ded{\Gamma}{a}{(\pro{x}{b}{c})} \quad \ded{\Gamma}{d}{b}}
  {\ded{\Gamma}{(\app{a}{d})}{c\sub{x}{d}}}$ \\

\\ (conv) & $\cfrac{\ded{\Gamma}{a}{b}
  \quad \ded{\Gamma}{b'}{p}}{\ded{\Gamma}{a}{b'}}$
  & $(p\in\starbox,\;$ $b\lrps{*}{\beta} b'$ or $b' \lrps{*}{\beta} b$\\
  && \hspace{14mm} or $b \lrps{*}{R} b'$ or $b' \lrps{*}{R} b)$ \\

 (prod) & $\cfrac{\ded{\Gamma}{a}{p} \quad \ded{\Gamma, x:a}{b}{q}}
  {\ded{\Gamma}{(\pro{x}{a}{b})}{q}}$
  & $(x\not\in\dom(\Gamma),\,p,q\in\starbox)$
\vspace{-2mm}
\end{tabular}
\end{figure}


\begin{definition}[Typing rules]
  A {\em declaration} is a pair $\decl{x}{a}$ made of a variable $x$
  and a term $a$.  An {\em environment} $\Gamma$ is a (possibly empty)
  ordered sequence of declarations of the form
  $\decl{x_1}{a_1},\ldots,\decl{x_n}{a_n}$, where all $x_i$ are
  distinct; $\dom(\Gamma)=\{x_1,\ldots,x_n\}$ is its domain,
  $\fv(\Gamma)=\bigcup_{x:a\in\Gamma}{\fv(a)}$ is its set of free
  variables, and $\Gamma(x_i)=a_i$.  A {\em typing judgement} is a
  triple $\ded{\Gamma}{a}{b}$ made of an environment $\Gamma$ and two
  terms $a,b$.  A term $a$ has {\em type $b$ in an environment}
  $\Gamma$ if the judgement $\ded{\Gamma}{a}{b}$ can be deduced by the
  rules of Figure \ref{fig-typ-rules}. An environment is {\em valid}
  if $\star$ can be typed in it. An environment is {\em algebraic} if
  every declaration has the form $\decl{x}{c}$, where $c$ is an
  algebraic type.
\end{definition}


The rules (sort), (cons) and (fun) are added to the rules of
CC~\cite{coquand88ic}. The (conv) rule expresses that types depend on
reductions via terms. In CC, the relation used in the side condition
is the monotonic, symmetric, reflexive, transitive closure of the
$\beta$-rewrite relation $(\lambda x\!:\!a . b)~c ~\lrps{}{\beta}~
b\sub{x}{c}$.

In our calculus, there are two kinds of computation rules: $\beta$-
(or proof-) reduction and the user-defined rewrite rules, denoted by
$\lrps{}{R}$. This contrasts with the other calculi of constructions,
in which the meaning of (conv) is fixed by the designer of the
language, while it depends on the user in our system. The unusual form
of the side condition of our conversion rule is due to the fact that
no proof of subject reduction is known for a conversion rule with the
more natural side condition $b\eqps{*}{\beta R} b'$.
See~\cite{barbanera97jfp} for details.


The structural properties of CC are also true in CAC.
See~\cite{barbanera97jfp} and~\cite{barendregt92jfp} for details.
We just recall the different term classes that compose the calculus.


\begin{definition}
  Let Kind be the set $\{K\in \term \barre \exists \Gamma,\,
  \ded{\Gamma}{K}{\Box}\}$ of {\em kinds}, Constr be the set $\{T\in
  \term \barre \exists \Gamma, \exists K\in \kind,\,
  \ded{\Gamma}{T}{K}\}$ of {\em type constructors}, Type be the set
  $\{\tau\in \term \barre \exists \Gamma,\,
  \ded{\Gamma}{\tau}{\star}\}$ of {\em types}, Obj be the set
  $\{u\in \term \barre \exists \Gamma, \exists \tau\in \type,\,
  \ded{\Gamma}{u}{\tau}\}$ of {\em objects}, and Thm be the set Constr
  $\cup$ Kind of {\em theorems}.
\end{definition}


\begin{lemma}
Kinds, type constructors and objects can be characterized as follows:
\begin{alist}{$\bullet$}
\item $K:=\star\barre\pro{x}{\tau}{K}\barre\pro{\alpha}{K}{K}$

\item $T:=\asort{s}\barre \alpha\barre \pro{x}{\tau}{\tau}\barre
  \pro{\alpha}{K}{\tau}
  \barre\abs{x}{\tau}{T}\barre\abs{\alpha}{K}{T}\barre(\app{T}{u})
  \barre(\app{T}{T})$

\item $u:=x\barre C(u_1,\ldots,u_n)\barre f(u_1,\ldots,u_n)\barre 
  \abs{x}{\tau}{u}\barre\abs{\alpha}{K}{u}\barre(\app{u}{u})\barre(\app{u}{T})$
\end{alist}
\noindent where $\alpha\in\varbox$ and $x\in\varstar$.
\end{lemma}


\subsection{Inductive types}

Inductive types have been introduced in CC for at least two reasons:
firstly, to ease the user's description of his/her specification by
avoiding the complicated impredicative encodings which were necessary
before; secondly, to transform inductive proofs into inductive
procedures via the Curry-Howard isomorphism. The logical consistency
of the calculus follows from the existence of a least fixpoint, a
property which is ensured syntactically in the Calculus of Inductive
Constructions by restricting oneself to strictly positive
types~\cite{coquand88colog}.


\begin{definition}[Positive and negative type positions]
Given an algebraic type $s$, its sets of positive and negative
positions are inductively defined as follows:

$\begin{array}{@{\hspace{-5mm}}c@{\hspace{3mm}}c}
Pos\!^+(s\in\sort)=\epsilon & Pos\!^-(s\in\sort)=\emptyset\\
Pos\!^+(s\tarrow t)=1\!\cdot\! Pos\!^-(s)\;\cup\; 2\!\cdot\! Pos\!^+(t) &
Pos\!^-(s\tarrow t)=1\!\cdot\! Pos\!^+(s)\;\cup\; 2\!\cdot\! Pos\!^-(t)\\
\end{array}$

Given an algebraic type $t$, we say that $s$ {\em does occur positively}
in $t$ if $s$ occurs in $t$, and each occurrence of $s$ in $t$ is at a
positive position.
\end{definition}


\begin{definition}[Inductive sorts]
Let $\asort{s}$ be a sort whose constructors are
$C_1,\ldots,$ $C_n$ and suppose that $C_i$ has type
$s^i_1\tarrow\ldots\tarrow s^i_{n_i}\tarrow \asort{s}$.
Then we say that:

(i) $\asort{s}$ is a {\em basic inductive sort}
  if each $s^i_j$ is $\asort{s}$ or a basic
  inductive sort 
smaller than $\asort{s}$ in $<_\sort$,

(ii) $\asort{s}$ is a {\em strictly positive inductive sort}
  if each $s^i_j$ is either a strictly
  positive inductive sort smaller than $\asort{s}$ in $<_\sort$,
  or of the form $s'_1\tarrow\ldots\tarrow s'_p\tarrow \asort{s}$ where each
  $s'_k$ is built from strictly positive inductive sorts
  smaller than $\asort{s}$ in $<_\sort$.

In the following, we will assume that every inductive sort of a user
specification is strictly positive.
\end{definition}


The sort $\nat$ whose constructors are $\onat:\nat$ and
$\snat:\nat \tarrow \nat$ is a basic sort.
The sort $\ord$ whose constructors are $\oord:\ord$, $\sord:\ord\tarrow \ord$
and $\limord:(\nat \tarrow \ord) \tarrow \ord$
is a strictly positive sort, since $\ord >_\sort \nat$.


\begin{definition}[Strictly positive recursors]
\label{def-str-pos-rec}
Let $\asort{s}$ be a strictly positive inductive sort generated by the
constructors $C_1,\ldots,C_n$ of respective types
$s^i_1\tarrow\ldots\tarrow s^i_{n_i}\tarrow \asort{s}$.
The associated
recursor $rec^\asort{s}_t$ of algebraic output type $t$ is a function
symbol of arity $n+1$, and type $\asort{s}\tarrow
t_1\tarrow\ldots\tarrow t_n\tarrow t$ where
$t_i=s^i_1\tarrow\ldots\tarrow s^i_{n_i}\tarrow
s^i_1\sub{\asort{s}}{t}\tarrow\ldots\tarrow
s^i_{n_i}\sub{\asort{s}}{t}\tarrow t$.  It is defined by the
rewrite rules:
\begin{center}
$rec^\asort{s}_t(C_i({\vec{a}}),\vec{b}) \;\arrow\;
   b_i\,\vec{a}\;\vec{d}~~$ where
\end{center}
$d_j=a_j$ if $\asort{s}$ is not in $s_i^j$, otherwise
$s^i_j=s'_1\ra\ldots\ra s'_p\ra \asort{s}$ and\\
$d_j=\abs{\vec{x}}{\vec{s'}\{\asort{s}\mapsto t\}}{ rec^\asort{s}_t
  (\app{a_j}{\vec{x}},\vec{b})}$.
\end{definition}


Via the Curry-Howard isomorphism, a recursor of a sort $\asort{s}$
corresponds to the structural induction principle associated to the
set of elements built from the constructors of $\asort{s}$.  Strictly
positive types are found in many proof assistants based on the
Curry-Howard isomorphism, e.g. in Coq~\cite{coq98}. Here are a few
recursors:

\vspace{-2mm}
\begin{center}
$\begin{array}{rcl@{\hspace{5mm}}rcl}
\boolrec(\true,u,v) &\arrow& u
&\natrec(\onat,u,v) &\arrow& u\\

\boolrec(\false,u,v) &\arrow& v
&\natrec(\snat(n),u,v) &\arrow& v~n~\natrec(n,u,v)\\
\end{array}$

\vspace{1mm}
$\begin{array}{rcl}
\ordrec(\oord,u,v,w) &\arrow& u\\
\ordrec(\sord(n),u,v,w) &\arrow&  v~n~\ordrec(n,u,v,w)\\
\ordrec(\limord(f),u,v,w) &\arrow& w~f~\lambda
n\!\!:\nat.\ordrec(f\,n,u,v,w)\\
\end{array}$
\end{center}

\vspace{-2mm}
\noindent $\boolrec$ is $\ifthenelse{t}$, and
$\natrec$ is G\"odel's higher-order primitive recursion operator.


\subsection{User-defined rules}


First, we define the syntax of terms that may be used for rewrite rules:

\begin{definition}[Rule terms]
Terms built up solely from constructors, function symbols and
variables of $\varstar$, are called
{\em algebraic}. Their set is defined by the following grammar:
\vspace{-2mm}
\begin{center}
$a:=x^\star\barre C(a_1,\ldots,a_n)\barre f(a_1,\ldots,a_n)$
\end{center}
\vspace{-1mm}
where $x^\star$ ranges over $\varstar$, $C$ over
$\const^n$ and $f$ over $\sig{}{n}$.  An algebraic term is {\em
  first-order} if its function symbols and constructors are
first-order, and {\em higher-order} otherwise. The set of {\em rule
  terms} is defined by the following grammar:
\vspace{-6mm}
\begin{center}
$a:=x^\star\barre \abs{x^\star}{s}{a}\barre (\app{a}{a})\barre
C(a_1,\ldots,a_n)\barre f(a_1,\ldots,a_n)$
\end{center}
\vspace{-1mm}
where $x^\star$ ranges over $\varstar$, $s$ over $\algtype$,
$C$ over $\const^n$
and $f$ over $\sig{}{n}$.
A rule term is {\em first-order} if it is a {\em first-order algebraic}
term, otherwise it is {\em higher-order}.
\end{definition}


\begin{definition}[Rewrite rules]
A {\em rewrite rule} is a pair $l\arrow r$ of rule terms such that
$l$ is headed by a function symbol $f$ which is said to be {\em
  defined}, and $\fv(r)\subseteq\fv(l)$.
Given a set $R$ of rewrite rules, a term $a$ rewrites to a term $b$ at
position $m\in\pos(a)$ with the rule $l\lrps{}{} r\in R$, written
$a\lrps{m}{R}b$ if $\subt{a}{m}=l\theta$ and $b=\remp{a}{r\theta}{m}$
for some substitution $\theta$.

A rewrite rule is {\em first-order} if $l$ and $r$ are both
first-order, otherwise it is {\em higher-order}. A first-order
rewrite rule $l\arrow r$ is {\em conservative} if no (free) variable
has more occurrences in $r$ than in $l$. The rules induce the
following quasi-ordering on function symbols: $f\ge_{\sig{}{}}g$ iff
$g$ occurs in a defining rule of $f$.
\end{definition}

We assume that first-order function
symbols are defined only by first-order rewrite rules.
Of course, it is always possible to treat a first-order function
symbol as an higher-order one. Here are examples of rules:


\vspace{-4mm}
\begin{center}
$\begin{array}{@{}r@{~}c@{~}l@{\hspace{-5mm}}r@{~}c@{~}l}
\ifthenelse{t}(\true,u,v) &\lrps{}{}& u
&\map{t,t'}(f,\nil{t}) &\lrps{}{}& \nil{t'}\\

\ifthenelse{t}(\false,u,v) &\lrps{}{}& v
&\map{t,t'}(f,\cons{t}(x,l)) &\lrps{}{}&
  \cons{t'}(\app{f}{x},\map{t,t'}(f,l))\\

\vspace{-2mm}\\

\plusnat(x,\onat) &\lrps{}{}& x
&\ack(\onat,y) &\lrps{}{}& \snat(y)\\

\plusnat(x,\snat(y)) &\lrps{}{}& \snat(\plusnat(x,y))
&\ack(\snat(x),\onat) &\lrps{}{}& \ack(x,\snat(\onat))\\

\plusnat(\plusnat(x,y),z) &\lrps{}{}& \plusnat(x,\plusnat(y,z))
&\ack(\snat(x),\snat(y)) &\lrps{}{}& \ack(x,\ack(\snat(x),y))\\
\end{array}$
\end{center}


Having rewrite rules in our calculus brings many benefits, in addition
to obtaining proofs in which computational steps are
transparent. In particular, it enhances the declarativeness of the
language, as examplified by the Ackermann's function, for which the
definition in Coq~\cite{coq98} must use two mutually recursive
functions.
For subject reduction, the
following properties will be needed:

\comment{
~\cite{coq98} is as follows:

{\small\tt \vspace{2mm}
\begin{tabular}{lllll}
\multicolumn{5}{l}{
Fixpoint ack[n:nat]:nat-$>$nat :=}\\
\quad&\multicolumn{4}{l}{
   Cases n of}\\
\quad&\quad&\multicolumn{3}{l}{
    O =$>$ [m:nat](S m)}\\
\quad&\quad&\multicolumn{3}{l}{
    $~|$ (S n') =$>$ Fix ack2 \{ack2/1:nat-$>$nat :=}\\
\quad&\quad&\quad&\multicolumn{2}{l}{
      [m:nat] Cases m of}\\
\quad&\quad&\quad&\quad&
        O =$>$ (ack n' (S O))\\
\quad&\quad&\quad&\quad&
        $~|$ (S m') =$>$ (ack n' (ack2 m'))\\
\quad&\quad&\quad&\multicolumn{2}{l}{
      end\}}\\
\quad&\multicolumn{4}{l}{
end.}
\end{tabular}
}
} 


\begin{definition}[Admissible rewrite rules]
\label{def-adm-rule}
A rewrite rule $l\arrow r$, where $l$ is headed by a function symbol
whose output type is $s$, is {\em admissible} if and only if it
satisfies the following conditions:

\begin{alist}{$\bullet$}
\item there exists an algebraic environment
$\Gamma_l$ in which $l$ is well-typed,
\item for any environment $\Gamma$,
$\ded{\Gamma}{l}{s}\imply\ded{\Gamma}{r}{s}$.
\end{alist}

\noindent We assume that rules use distinct variables and note by $\gamR$
the union of the $\Gamma_l$'s.
\end{definition}

\comment{In comparison to \cite{barbanera94lics} where the
  decidability of similar conditions called {\em cube-embeddability}
  is proved only for algebraic terms, we can prove it for a larger
  class of rule terms including abstractions and applied variables.
As an example, let us prove the admissibility of the following rewrite
rule: $\plusnat(u,\snat(v))$ $\lrps{}{}$ $\snat(\plusnat(u,v))$. 
Well-typedness condition is satisfied with
$\Gamma=u:\nat,v:\nat$.
\comment{Algebraicity follows from the fact that
each variable is the argument of a function symbol.}
Type preservation follows by the same token.}


\subsection{Definition of the General Schema}

\comment{
We are now ready for describing a schema for the higher-order rules,
that the user definitions should follow. This schema is inspired
from~\cite{jouannaud97tcs}, although the formulation is quite
different, and the expressiveness slightly enhanced.

From now on, we assume that each (defined) higher-order function
symbol $f\in\sig{s_1,\ldots,s_n,t}{}$ comes along with a status,
lexicographic or multiset, $stat_f\in\{lex,mul\}$
(see~\cite{dershowitz90book} for more details on lexicographic and
multiset comparisons). A more elaborated notion of status will be used
later. We also assume that the ordering $>_{\cal F}$ is well-founded
on higher-order function symbols, therefore ruling out mutual
recursions. This is not a real restriction, since our schema is
powerful enough to encode mutual recursions via product types, which
can themselves be expressed as inductive types. We now define a key
set of terms.
}


Let us consider the example of a strictly positive recursor
rule, for the sort $\ord$:
\begin{center}
$\ordrec(\limord(f),u,v,w) \arrow
  w~ f~ \lambda n\!\!:\!\!\nat.\ordrec(f\,n,u,v,w)$
\end{center}
To prove the decreasingness of the recursive call arguments, one would
like to compare $\limord(f)$ with $f$, and not $\limord(f)$ with
$(\app{f}{n})$.  To this end, we introduce the notion of the {\em
  critical subterm} of an application, and then interpret a function
call by the critical subterms of its arguments. Here, $f$ will be the
critical subterm of $(\app{f}{n})$, hence resulting in the desired
comparison.


\begin{definition}[$\gams$-critical subterm]
\label{csd}
Given an algebraic type $s$ and an environment $\Gamma$, a term $a$ is
a {\em $\gams$-term} if it is typable in $\Gamma$ by an algebraic type
in which $s$ occurs positively. A term $b$ is a {\em $\gams$-subterm}
of a term $a$, $a\supertermeqgams{\Gamma}{s} b$, if $b$ is a subterm
of $a$, of which each superterm is a $\gams$-term.
Writing a $\gams$-term $a$ in its {\em application form} $a_1\ldots
a_n$, where $a_1$ is not an application, its {\em $\gams$-critical
  subterm} $\crit{\Gamma}{s}(a)$ is the smallest $\gams$-subterm
$a_1\ldots a_k$ (see Figure~\ref{figure}).
\end{definition}


For a higher-order function symbol, the arguments that have to be
compared via their critical subterm, are said to be at {\em inductive
  positions}. They correspond to the arguments on which the function
is inductively defined. Next, we define a notion of status that allows
users to precise how to compare the arguments of recursive calls.
Roughly speaking, it is a simple combination of multiset and
lexicographic comparisons.
 

\begin{definition}[Status orderings]
\label{statusd}
A {\em status} of arity $n$ is a term of the form
$lex(t_1,\ldots,t_p)$ where $t_i$ is either $x_j$ for some
$j\in[1..n]$, or a term of the form $mul(x_{k_1},\ldots,x_{k_q})$ such
that each variable $x_i$, $1\leq i \leq n$, occurs at most once.  A
position $i$ is {\em lexicographic} if there exists $j$ such that
$t_j=x_i$.  A {\em status term} is a status whose variables are
substituted by arbitrary terms of CAC.

Let $stat$ be a status of arity $n$, $I$ be a subset of the
lexicographic positions of $stat$, called {\em inductive positions},
$S=\{>^i\}_{i\in I}$ a set of orders on terms indexed by $I$, and $>$
an order on terms.  We define the corresponding {\em status ordering},
$>_{stat}^S$ on sequences of terms as follows:

\begin{blist}{$\bullet$}
\item $(a_1,\ldots,a_n) >_{stat}^S (b_1,\ldots,b_n)$
  iff $stat\subvec{a} >_{stat}^S stat\subvec{b}$,
\item $lex(c_1,\!\ldots\!,c_p) >_{lex(t_1,\ldots,t_p)}^S
  lex(d_1,\!\ldots\!,d_p)$ iff $(c_1,\!\ldots\!,c_p)
  \lexx{(>_{t_1}^S,\!\ldots\!,>_{t_p}^S)} (d_1,\!\ldots\!,d_p)$,
\item $>_{x_i}^S$ is $>^i$ if $i\in I$, otherwise it is $>$,
\item $mul(c_1,\ldots,c_q) >_{mul(x_{k_1},\ldots,x_{k_q})}
    mul(d_1,\ldots,d_q)$ iff $\{c_1,\ldots,c_q\} \mul{>}
    \{d_1,\ldots,d_q\}$.
\end{blist}

\noindent Note that it boils down to the usual lexicographic ordering if
$stat=lex(x_1,\ldots,x_n)$ or to the multiset ordering if
$stat=lex(mul(x_1,\ldots,x_n))$. $>_{stat}^S$ is well-founded if so
is $>$ and each $>^i$.
\end{definition}

For example, let $>$ and $\succ$ be some orders, $stat= lex(x_2,
mul(x_1, x_3))$, $I=\{1\}$, and $S=(\succ)$. Then, $(a_1,a_2,a_3)
>_{stat}^S (b_1,b_2,b_3)$ iff $a_2 \succ b_2$, or else $a_2= b_2$ and
$\{a_1,a_3\} >_{mul} \{b_1,b_3\}$.


\begin{definition}[Critical interpretation]
  Given an environment $\Gamma$, the {\em critical interpretation
    function} $\critint{}$ of a function symbol $f\in
  \sig{s_1,\ldots,s_n,s}{}$ is:
\begin{alist}{$\bullet$}
\item $\critint{}(a_1,\ldots,a_n)=
  (\critint{1}(a_1),\ldots,\critint{n}(a_n))$,
\item $\critint{i}(a_i)=a_i$ if $i\not\in\ind(f)$,
\item $\critint{i}(a_i)=\crit{\Gamma}{s_i}(a_i)$ if $i\in\ind(f)$.
\end{alist}
\noindent The {\em critical ordering} associated to $f$ is
$>_{f,_\Gamma}= \superterm\!\!_{stat_f}^S$, where $S=
(\supertermgams{\Gamma}{s_i})_{i\in\ind(f)}$.
\end{definition}

According to Definition~\ref{statusd}, the critical ordering is
nothing but the usual subterm ordering at non-inductive positions, and
the critical subterm ordering of Definition~\ref{csd} at inductive
positions.


\comment{
\begin{figure}[ht]
\begin{center} 
\raisebox{2.5cm}{\scalebox{.5}{\includegraphics{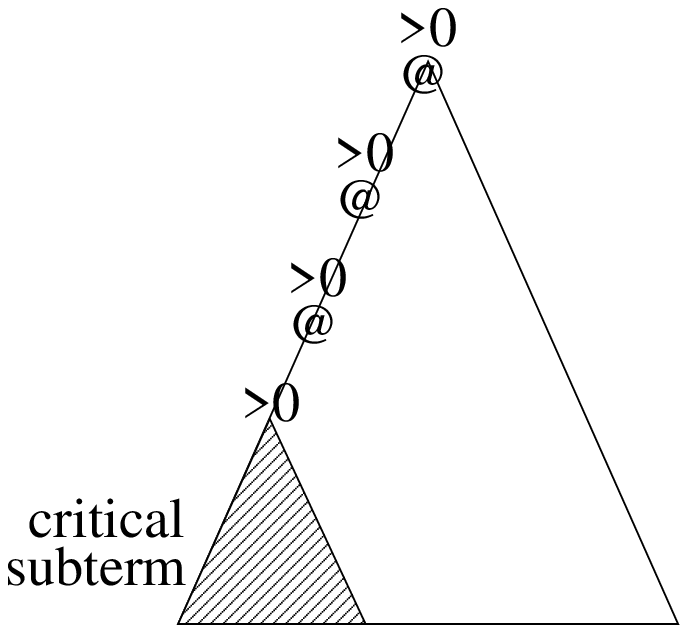}}}
\hspace{1cm}
\scalebox{.5}[.4]{\includegraphics{rec-pos.ps}}

\quad
(a) Critical subterm
\quad\quad
(b) Extended critical ordering
\end{center}
\caption{Extensions of the General Schema\label{figure}}
\end{figure}
}

\begin{figure}[ht]
\vspace{-2mm}
\begin{center} 
\scalebox{.4}{\includegraphics{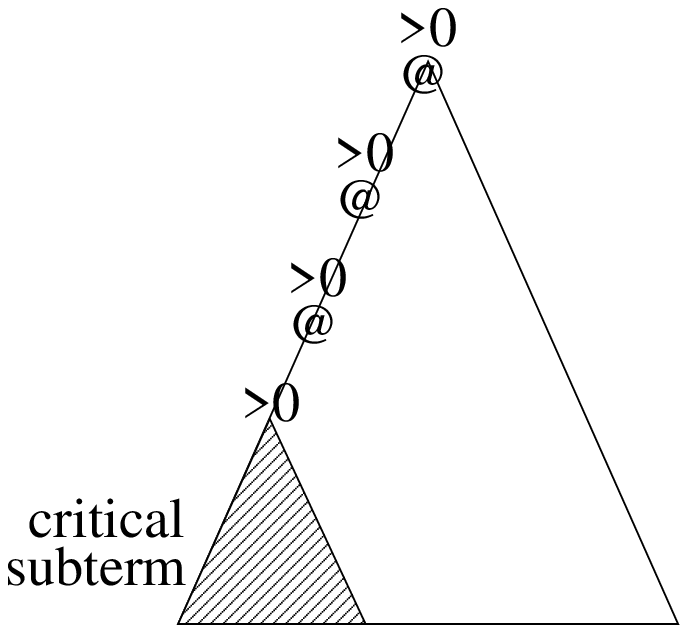}}
\end{center}
\vspace{-5mm}\caption{Critical subterm\label{figure}}
\vspace{-4mm}
\end{figure}

We are now ready for describing the schema for higher-order rewrite
rules. Given some lefthand side rule, we define a set of acceptable
righthand sides, called computable closure. In the next section, we
prove that it preserves strong normalization.


\begin{definition}[Accessible subterms]
  A term $b$ is said to be {\em accessible} in a well-typed term $c$
  if it is a subterm of $c$ which is typable by a basic inductive
  sort, or if there exists $p\in\Pos{c}$ such that $c|_p=b$, and $\forall
  q < p$, $c|_q$ is headed by a constructor. $b$ is said to be {\em
    accessible} in $\vec{c}$ if it is so in some $c\in\vec{c}$.
\end{definition}


\begin{definition}[Computable closure]
\label{ccd}
Given an algebraic environment $\Gamma$ containing $\gamR$ and a term
$f(\vec{c})$ typable in $\Gamma$, the {\em computable closure}
$\CS{f,\Gamma}{\vec{c}}$ of $f(\vec{c})$ in $\Gamma$ is defined as the
least set of $\Gamma$-terms containing all terms accessible in
$\vec{c}$, all variables in $\dom(\Gamma) \setminus \fv(\vec{c})$, and
closed under the following operations:

\begin{list}{}{\parsep=0ex\topsep=0ex\itemsep=0ex\leftmargin=5mm
  \itemindent=-5mm}
\item (i) constructor application: let $C$ be a constructor of type
  $s_1\ra\ldots\ra s_n\ra \asort{s}$; then
  $C(\vec{u})\in\CS{f,\Gamma}{\vec{c}}$ iff
  $u_i:s_i\in\CS{f,\Gamma}{\vec{c}}$ for all $i\in[1..n]$,
  
\item (ii) defined application: let $g\in\sig{s_1,\ldots,s_n,t}{}$
  such that $g\ltF f$; then $g(\vec{u})\in \CS{f,\Gamma}{\vec{c}}$ iff
  $u_i:s_i\in\CS{f,\Gamma}{\vec{c}}$ for all $i\in[1..n]$,
  
\item (iii) application: let $u \!:\! s\!\ra\!
  t\in \!\CS{f,\Gamma}{\vec{c}}$ and $v\!:\!s \in \!\CS{f,\Gamma}{\vec{c}}$;
  then $(u v)\in \!\CS{f,\Gamma}{\vec{c}}$,

\item (iv) abstraction: let $u\in\CS{f,\Gamma}{\vec{c}}$ and $x:s\in\Gamma$;
  then $\lambda x\!:\!s . u \in\CS{f,\Gamma}{\vec{c}}$,
  
\item (v) reduction: let $u \in \CS{f,\Gamma}{\vec{c}}$, and $v$ be a
  reduct of $u$ using a $\beta$-rewrite step or a higher-order rewrite
  rule for a function symbol $g\ltF f$; then $v\in
  \CS{f,\Gamma}{\vec{c}}$,
  
\item (vi) recursive call: let $\vec{c'}$ be a vector of $n$ terms in
$\CS{f,_\Gamma}{\vec{c}}$ of respective types $s_1,\ldots,s_n$, such
that $\critint{}(\vec{c})= \vec{c} ~~>_{f,_\Gamma}~
\critint{}(\vec{c'})$; then $f(\vec{c'})\in \CS{f,_\Gamma}{\vec{c}}$.
\end{list}
\end{definition}

A useful finite approximation of this infinite set is defined by the
Coquand's notion of {\em structurally smaller}~\cite{coquand92types},
where only cases (i), (iii), (v) (one $\beta$-step only) and (vi) are
used, with a multiset status which forbids the use of nested
recursions. Our definition is therefore richer for two independent
reasons. Note further that Coquand restricts himself to the cases for
which his ordering is well-founded, a property that we think related
to the positivity condition.

This can also be compared with the current criterion used in Coq for
accepting function definitions by fixpoint and constructor
matching~\cite{cornes97these}. Functions are defined by induction on
one argument at a time, this argument must be constructor headed, and
recursive calls can be made only with its immediate subterms.
We are now ready for defining the schema:


\begin{definition}[General Schema]
\label{general-schema}
\label{gs}
A set $R$ of rewrite rules satisfies the General Schema if

(i) its first-order part is conservative and strongly normalizing,

(ii) each higher-order function $f\in{\cal F}_{s_1, \ldots, s_n, t}$
is defined by a set of admissible rewrite rules of the form
$f(\vec{c})\lrps{}{} e$ such that $e\in\CS{f,\Gamma}{\vec{c}}$ for
some algebraic $\Gamma$ containing $\gamR$ (the environment in which
the rules of $R$ are defined).
\end{definition}

All pattern-matching definitions given so far satisfy the General
Schema, including the first-order ones. We could have imposed that the
first-order rules also satisfy the General Schema: this would have
simplified our definition, but at the price of restricting the
expressivity for the first-order rules.  In our formulation, the
strong normalization property of the first-order rules has to be
proved beforehand. Tools exist that do the job automatically for many
practical examples.
Note that recursor rules of any strictly positive inductive type
satisfy the General Schema:


\begin{lemma}
  The recursor rules for strictly positive inductive sorts satisfy the
  General Schema.
\end{lemma}

\comment{
\begin{proof}
  Let $\asort{s}$ be a strictly positive inductive sort and $t$ be an
  arbitrary type.  We prove that the righthand side
  $(b_i~\vec{a}~\vec{b'})$ belong to
  $\CS{rec^\asort{s}_t,_\Gamma}{C_i({\vec{a}}),\vec{b}}$. Let us take
  $\ind(rec^\asort{s}_t)= \{1\}$ and $stat_{rec^\asort{s}_t}=
  lex(x_1)$.  The first argument of a recursive call is of the form
  $(a_j~\vec{x})$ where $\vec{x}$ are bound variables. Hence,
  $\crit{\gamR}{\asort{s}}(a_j~\vec{x})= a_j$ and $C_i(\vec{a})
  \supertermgams{\gamR}{\asort{s}} a_j$ since $C_i(\vec{a})$ has type
  $\asort{s}$ and $a_j$ has type $s^i_1\ra \ldots\ra s^i_{n_i}\ra
  \asort{s}$ where $s^i_k$ are built from strictly positive inductive
  sorts smaller than $\asort{s}$ in $>_\sort$.\qed\\
\end{proof}
} 


\subsection{CAC computations}


\begin{definition}[Reduction relation]
  Given a set $R$ of rewrite rules satisfying the General Schema,
  including the set $Rec$ of recursor rules of a given user
  specification, the CAC rewrite relation is $\lrps{}{}\; =
  \lrps{}{\beta} \cup \lrps{}{R}$. The CAC reduction relation is its
  reflexive and transitive closure denoted by $\lrps{*}{}$. Its
  transitive closure is denoted by $\lrps{+}{}$. Its reflexive,
  symmetric and transitive closure is denoted by $\eqps{*}{}$.
  A term is in {\em normal form} if it cannot be
  $\beta$-reduced, $Rec$-reduced or $R$-reduced. An {\em expansion} is
  the inverse of a reduction: $a$ expanses to $b$ if $b$ reduces to
  $a$.
\end{definition}

Our calculus enjoys the subject reduction property, that is,
preservation of types under reductions. The proof uses a weak version
of confluence, see~\cite{barbanera97jfp}.

Full confluence is proved after strong-normalization, by using
Newman's Lemma, and by assuming there are no critical pairs between
any two higher-order rules, and between the higher-order rules, the
first-order rules and the $\beta$-reduction rule (by considering that
the abstraction is an unary function symbol, and the application a
binary one).


%% file: strong-normalization.tex
\section{Strong-normalization}
\label{strong-normalization}

A term is {\em strongly normalizable} if any reduction issuing from it
terminates. Strong-normalization and confluence together imply the
logical soundness of the system as well as the decidability of
type-checking. In this section, we investigate only the former. Let
$\sn$ be the set of strongly normalizable terms.


To prove the strong normalization property for well-typed terms, we
use the well known proof technique of Girard dubbed ``reducibility
candidates''~\cite{girard88book}, further extended by Coquand and
Gallier to the Calculus of Constructions~\cite{coquand90lf}. Note that
these proofs use well-typed candidates, that is, sets of well-typed
terms. There exists proofs with lighter notations based on untyped
candidates~\cite{geuvers94types}, but which do not allow one to reason
about the type of the elements of a reducibility candidate, as it will
be necessary to do with our extension of the General Schema. For a
comprehensive survey of the method, see~\cite{gallier90book}.

\comment{
The idea behind that proof technique is the following. A proof of
$\ded{\Gamma}{a}{b}\imply a\in\sn$ by induction on the structure of
$a$ does not go through because of the application case: if $a$ and
$b$ are strongly normalizable terms, the application $(a~b)$ may not
be strongly normalizable. Tait's idea, later reformulated by Girard in
a more general context, is to use a stronger induction
hypothesis by proving $\ded{\Gamma}{a}{b}\imply a\in\inter{b}$, for
some interpretation function $\inter{~}$ ``stable by application''
such that $\inter{b}\subseteq\sn$. For a comprehensive survey of the
method, see~\cite{gallier90book}.
} 

The strong normalization proof of Coquand and Gallier can easily be
tailored to our need. It suffices to define an adequate interpretation
for the inductive types, and to prove that, if the arguments of a
function call belong to the interpretation of their type, then the
function call itself belongs to the interpretation of its output type.
We recall the definitions that are necessary for the understanding of
our extension, and refer the reader to~\cite{coquand90lf} for a
complete exposition.
 
 
\subsection{Interpretation of theorems}

\newcommand{\dd}[2]{#1\!\these\! #2}
\newcommand{\cc}[1]{{\cal C}_{#1}}
\newcommand{\ty}[1]{{\cal T}_{#1}}
\newcommand{\nn}[1]{\sn_{#1}}
\newcommand{\can}[1]{can_{#1}}


\begin{definition}[Reducibility candidates]
\label{def-red-cand}
We define the set $\neutr$ of {\em neutral terms} as being the set of
terms that are not an abstraction or constructor headed. Let
$\ty{\Delta,A}= \{\dd{\Delta'}{a} \barre \ded{\Delta'}{a}{A},
~\Delta'\supseteq \Delta\}$, $\nn{\Delta,A}= \{\dd{\Delta'}{a}\in
\ty{\Delta,A} \barre a\in\sn\}$.

Given a valid environment $\Delta$, the family $\cc{}$ of saturated
sets $\cc{\Delta,A}$ where $A$ is a $\Delta$-theorem, is
defined by the properties listed below.

\begin{blist}{}
\item 1. If $A=\Box$, then $\cc{\Delta,A}$ is the set
  $\{\nn{\Delta,\Box}\}$.
  
\item 2. If $A$ is a $\Delta$-type or a $\Delta$-kind, then
  $\cc{\Delta,A}$ is the set of non empty sets $S\subseteq
  \nn{\Delta,A}$ such that the following properties hold:

  \begin{blist}{}
  \item (S1) $S\supseteq \{\dd{\Delta'}{x\vec{a}}\in \ty{\Delta,A} \barre
    x\in\var$ and $\vec{a}\in \sn\}$.
  \item (S2) For every neutral term $t$ such that $\dd{\Delta'}{t}\in
    \ty{\Delta,A}$, if, for every immediate reduct $t'$ of $t$,
    $\dd{\Delta'}{t'}\in S$, then $\dd{\Delta'}{t}\in S$.
  \item (S3) Whenever $\dd{\Delta'}{t}\in S$ and
    $\Delta'\subseteq\Delta''$, then $\dd{\Delta''}{t}\in S$.
  \item (S4) Whenever $\dd{\Delta'}{t}\in S$ and $t'$ is a reduct
    of $t$, then $\dd{\Delta'}{t'}\in S$.
  \end{blist}

\item 3. If $A$ is a type constructor of type $\pro{x}{B}{C}$ in
  $\Delta$, then $\cc{\Delta,A}$ is the set of functions with the
  following properties:

  \begin{blist}{}
  \item (a) If $B$ is a kind, then
    \begin{blist}{}
    \item $\bullet$ $f\in \cc{\Delta,A}$ is a function with domain
      $\{(\dd{\Delta'}{T},S) \barre \dd{\Delta'}{T}\in
      \ty{\Delta,B}$ and $S\in \cc{\Delta',T}\}$ such that
      $f(\dd{\Delta'}{T},S)\in \cc{\Delta',AT}$,
    \item $\bullet$ $f(\dd{\Delta'}{T_1},S_1) =
      f(\dd{\Delta'}{T_2},S_2)$ whenever $T_1\eqps{*}{} T_2$.
    \end{blist}

  \item (b) If $B$ is a type, then
    \begin{blist}{}
    \item $\bullet$ $f\in \cc{\Delta,A}$ is a function with domain
      $\ty{\Delta,B}$ such that $f(\dd{\Delta'}{t})\in
      \cc{\Delta',At}$,
    \item $\bullet$ $f(\dd{\Delta'}{t_1}) = f(\dd{\Delta'}{t_2})$
      whenever $t_1\eqps{*}{} t_2$.
    \end{blist}
  \end{blist}

\end{blist}
\end{definition}

\noindent Compared to~\cite{coquand90lf}, we extended (S2) to neutral
terms to take care of functions, and added (S4) to insure that
reducibility candidates are stable by reduction.

\comment{

\begin{lemma}[Non-emptiness]
Saturated sets are not empty.
\end{lemma}

\begin{proof}
  Our new conditions do not call into question the proof
  of~\cite{coquand90lf}. We recall it since it defines some canonical
  saturated sets which will be useful later. The proof is by induction
  on the following complexity measure $c$ on theorems, which is
  compatible with the conversion relation: $c(\Box)=0$, $c(T)=0$ if
  $T$ is a type, $c(\star)=1$, and $c(\pro{x}{A}{K})= 1+
  max(c(A),c(K))$.
  
  We define $\can{\Delta,\Box}= \nn{\Delta,\Box}$ and
  $\can{\Delta,\star}= \nn{\Delta,\star}$. Let $A$ be a type
  constructor valid in an environment $\Delta$. If $A$ is a type, then
  it is easy to verify that $\can{\Delta,A}= \nn{\Delta,A}\in
  \cc{\Delta,A}$. If $A$ is of type $\pro{x}{B}{C}$ in $\Delta$, then
  we define $\can{\Delta,A}$ as being the function such that
  $\can{\Delta,A}(\dd{\Delta'}{T},S)= \can{\Delta',AT}$ if $B$ is a
  kind, or $\can{\Delta,A}(\dd{\Delta'}{t})= \can{\Delta',At}$ if $B$
  is a type.
\end{proof}
} 


\begin{definition}[Interpretation of algebraic types]
\label{lem-int-alg-types}
Given a valid environment $\Delta$, we define the interpretation of
algebraic types as follows:
\begin{alist}{$\bullet$}
\item $\can{\Delta,\asort{s}}= \{\dd{\Delta'}{a}\in
  \nn{\Delta,\asort{s}} \barre$ if $a\arrowstar C(\vec{b})$ and
  $\tau(C)= s_1 \!\ra\! \ldots \!\ra\! s_n \!\ra\! \asort{s}$, then
  $\dd{\Delta'}{b_i}\in \can{\Delta,s_i}$ for every $i\in [1..n]\}$,
\item $\can{\Delta,s\ra t}= \{\dd{\Delta'}{a}\in \ty{\Delta,s\ra t}
  \barre \forall ~\Delta''\subseteq \Delta', ~\forall
  ~\dd{\Delta''}{b}\in \can{\Delta,s},~$\\ $\dd{\Delta''}{ab}\in
  \can{\Delta,t}\}$.
\end{alist}
\end{definition}

Let us justify the definition. Since $>_\sort$ is assumed to be
well-founded, our hypothesis is that the definition makes sense for
every algebraic type built from sorts strictly smaller than a given
sort $\asort{s}$. Let $P$ be the set of subsets of
$\nn{\Delta,\asort{s}}$ that contains all strongly normalizable terms
that do not reduce to a term headed by a constructor of $\asort{s}$.
$P$ is a complete lattice for set inclusion.  Given an element $X\in
P$, we define the following function on algebraic types built from
sorts smaller than $\asort{s}$: $R_X(\asort{s})= X$, $R_X(\asort{t})=
\can{\Delta,\asort{t}}$ and $R_X(s\ra t)= \can{\Delta,s\ra t}$. Now,
let $F:P\rightarrow P,\,X\mapsto X\cup Y$ where
$Y=\{a\in\nn{\Delta,\asort{s}} \barre$ if $a\arrowstar\,C(\vec{b})$
and $\tau(C)= s_1\ra \ldots \ra s_n\ra \asort{s}$ then $b_i\in
R_X(s_i)$ for every $i\in [1..n]\}$.  Since inductive sorts are
assumed to be positive, one can show that $F$ is monotone. Hence, from
Tarski's Theorem, it has a least fixed point
$\can{\Delta,\asort{s}}\in \cc{\Delta,\asort{s}}$.


\begin{definition}[Well-typed substitutions]
Given two valid environments $\Delta$ and $\Gamma$, a substitution
$\theta$ is a {\em well-typed substitution from $\Gamma$ to $\Delta$}
if $dom(\theta)\subseteq \dom(\Gamma)$ and, for every variable
$x\in\dom(\Gamma)$, $\ded{\Delta}{x\theta}{\Gamma(x)\theta}$.
\end{definition}

\comment{
Note that this definition is correct since, in a valid environment
$x_1:a_1,\ldots,x_n:a_n$, $\fv(a_i)\subseteq \{x_1,\ldots, x_{i-1}\}$
for every $i\in [1\ldots n]$.
} 


\begin{definition}[Candidate assignments]
  Given two valid environments $\Delta$ and $\Gamma$, and a well-typed
  substitution $\theta$ from $\Gamma$ to $\Delta$, a {\em candidate
    assignment compatible with $\theta$} is a function $\xi$ from
  $\varbox$ to the set of saturated sets such that, for every variable
  $\alpha\in\dom(\Gamma)\cap \varbox$, $\xi(\alpha)\in
  \cc{\Delta,\alpha\theta}$.
\end{definition}

Compared to~\cite{coquand90lf} where well-typed substitutions and
candidate assignments are packaged together, we prefer to
separate them since the former is introduced to deal with
abstractions, while the latter is introduced to deal with
polymorphism. We are now ready to give the definition of the
interpretation of theorems.


\renewcommand{\int}[5]{[\![\dd{#1}{#2}]\!]_{#3,#4,#5}}
\newcommand{\inta}[1]{\int{\Gamma}{#1}{\Delta}{\theta}{\xi}}

\begin{definition}[Interpretation of theorems]
  Given two valid environments $\Delta$ and $\Gamma$, a well-typed
  substitution $\theta$ from $\Gamma$ to $\Delta$, and a candidate
  assignment $\xi$ compatible with $\theta$, we define the
  interpretation of $\Gamma$-theorems as follows:

\begin{list}{$\bullet$}{\parsep=1mm\topsep=0mm\itemsep=0mm}
\item $\inta{\Box} = \nn{\Delta,\Box}$,

\item $\inta{\star} = \nn{\Delta,\star}$,

\item $\inta{\asort{s}} = \can{\Delta,\asort{s}}$,

\item $\inta{\alpha} = \xi(\alpha)$,
  
\item $\inta{\abs{x}{\tau}{T}}=$ the function which associates
  $\int{\declb{\Gamma}{x}{\tau}}{T}{\Delta'}{\theta \{x\mapsto
    t\}}{\xi}$\\ to every $\dd{\Delta'}{t}\in \ty{\Delta,\tau\theta}$,
  
\item $\inta{\abs{\alpha}{K'}{T}}=$ the function which
    associates\\ $\int{\declb{\Gamma}{\alpha}{K'}}{T}{\Delta'}{\theta
    \{\alpha\mapsto T'\}}{\xi \{\alpha\mapsto S\}}$\\ to every
  $(\dd{\Delta'}{T'},S)\in \{(\dd{\Delta'}{T'},S) \barre
  \ded{\Delta'}{T'}{K'\theta}, ~\Delta'\supseteq \Delta, ~S\in
  \cc{\Delta',T'}\}$,

\item $\inta{T~t} = \inta{T}(\dd{\Delta}{t\theta})$

\item $\inta{T~T'} = \inta{T}(\dd{\Delta}{T'\theta},\inta{T'})$
  
\item $\inta{\pro{x}{\tau}{A}} = \{\dd{\Delta'}{a}\in
  \ty{\Delta,\Pi x:\tau\theta.A\theta} \barre \forall
  \Delta''\supseteq \Delta',~ \forall ~\dd{\Delta''}{t}\in\\
  \int{\Gamma}{\tau}{\Delta''}{\theta}{\xi}, ~\dd{\Delta''}{at}\in
  \int{\declb{\Gamma}{x}{\tau}}{A}{\Delta''}{\theta\{x\mapsto
    t\}}{\xi}\}$,
  
\item $\inta{\pro{\alpha}{K}{A}} = \{\dd{\Delta'}{a}\in
  \ty{\Delta,\Pi \alpha:K\theta.A\theta} \barre \forall
  \Delta''\supseteq \Delta',~ \forall ~\dd{\Delta''}{T}\in
  \int{\Gamma}{K}{\Delta''}{\theta}{\xi}, ~\forall S\in
  \cc{\Delta'',T}, ~\dd{\Delta''}{aT}\in
  \int{\declb{\Gamma}{\alpha}{K}}{A}{\Delta''}{\theta\{\alpha\mapsto
    T\}}{\xi\{\alpha\mapsto S\}}\}$.
\end{list}
\end{definition}


The last two cases correspond to the ``stability by application''. The
well-definedness of this definition is insured by the following lemma.

\begin{lemma}[Interpretation correctness]
\label{correct}
Assume that $\Delta$ and $\Gamma$ are two valid environments, $\theta$
is a well-typed substitution from $\Gamma$ to $\Delta$, and $\xi$ is a
candidate assignment compatible with $\theta$. Then, for every
$\Gamma$-theorem $A$, $\inta{A}\in \cc{\Delta,A\theta}$.
\end{lemma}



\noindent We are now able to state the main lemma
for the strong normalization theorem.

\begin{definition}[Reducible substitutions]
  Given two valid environments $\Delta$ and $\Gamma$, a well-typed
  substitution $\theta$ from $\Gamma$ to $\Delta$, and a candidate
  assignment $\xi$ compatible with $\theta$, $\theta$ is said to be
  {\em valid with respect to $\xi$} if, for every variable $x\in
  \dom(\Gamma)$, $\dd{\Delta}{x\theta}\in \inta{\Gamma(x)}$.
\end{definition}

\begin{lemma}[Main lemma]
\label{lem-main-for-sn}
Assume that $\ded{\Gamma}{a}{b}$, $\Delta$ is a valid environment,
$\theta$ is a well-typed substitution from $\Gamma$ to $\Delta$, and
$\xi$ is a candidate assignment compatible with $\theta$. If $\theta$
is valid with respect to $\xi$, then $\dd{\Delta}{a\theta}\in \inta{b}$.
\end{lemma}

\begin{proof}
  As in~\cite{coquand90lf}, by induction on the structure of the
  derivation. We give only the additional cases. The case (cons) is
  straightforward. The case (fun) is proved by
  Theorem~\ref{th-red-omega} to come for the case of higher-order
  function symbols, and by~\cite{jouannaud97tcs} for the case of
  first-order function symbols.\qed
\end{proof}


\begin{theorem}[Strong normalization]
\label{th-sn}
Assume that the higher-order rules satisfy the General Schema.
Then, any well-typed term is strongly normalizable.
\end{theorem}

\begin{proof}
Application of the Main Lemma, see~\cite{coquand90lf} for details.
\end{proof}

\comment{
\begin{proof}
  It suffices to take $\Delta=\Gamma$, the identity substitution for
  $\theta$, and for $\xi$, the candidate assignment such that,
  $\forall \alpha\!\in\! \dom(\Gamma)\cap \varbox, ~\xi(\alpha)=
  \can{\Gamma,\Gamma(\alpha)}$, which is compatible with the identity
  substitution.\qed
\end{proof}
} 


\subsection{Reducibility of higher-order function symbols}


One can see that the critical interpretation is not compatible with
the reduction relation, and not stable by substitution either.  We
solve this problem by using yet another interpretation function for
terms enjoying both properties and relating to the previous one as
follows:

\begin{definition}[Admissible recursive call interpretation]
  A {\em recursive call interpretation} for a function symbol $f$ is
  given by:

\begin{blist}{}
\item (i) a function $\rcint{\Gamma}{}$ operating on
  arguments of $f$, for each environment $\Gamma$,
\item (ii) a status ordering $\ge_{stat_f}^S$ where $S$ is a set
  of orders indexed by $\ind(f)$.
\end{blist}

\noindent A recursive call interpretation is {\em admissible} if it
satisfies the following properties:

\begin{list}{$\bullet$}{\parsep=0ex\topsep=0ex\itemsep=0ex
                        \leftmargin=2mm\itemindent=0mm}
\item[\bf (Stability)] Assume that $f(\vec{c'})\in
  \CS{f,_\Gamma}{\vec{c}}$, hence $\critint{}(\vec{c}) = \vec{c}
  >_{f,_\Gamma} \critint{}(\vec{c'})$, $\Delta$ is a valid
  environment, and $\theta$ is a well-typed substitution from $\Gamma$
  to $\Delta$ such that $\vec{c}\theta$ are strongly normalizable
  terms. Then, $\rcint{\Delta}{}(\vec{c}\theta) >_{stat_f}^S
  \rcint{\Delta}{}(\vec{c'}\theta)$.
  
\item[\bf (Compatibility)] Assume that $s$ is the output type of $f$,
  $\vec{a}$ and $\vec{a'}$ are two sequences of strongly normalizable
  terms such that $\ded{\Delta}{f(\vec{a})}{s}$ and $\vec{a}\arrowstar
  \vec{a'}$. Then, $\rcint{\Delta}{}(\vec{a}) \ge_{stat_f}^S
  \rcint{\Delta}{}(\vec{a'})$.
\end{list}
\end{definition}


The definition of the actual interpretation function, which is
intricate, can be found in the full version of the paper.
Before to prove the reducibility of higher-order function symbols, we
need the following result.


\begin{lemma}[Compatibility of accessibility with reducibility]
\label{lem-comp-acc-red}
\hfill\\
If $\dd{\Delta}{a}\in \can{\Delta,A}$ and $b\in \ty{\Delta,B}$ is
accessible in $a$, then $\dd{\Delta}{b}\in \can{\Delta,B}$.
\end{lemma}



\begin{theorem}[Reducibility of higher-order function symbols]
\label{th-red-omega}
\hfill\\
Assume that the higher-order rules satisfy the General Schema. Then,
for every higher-order function symbol $f\in \sig{s_1,\ldots,
  s_n,s}{}$, $\dd{\Delta}{f(\vec{a})}\in \can{\Delta,s}$ provided that
$\ded{\Delta}{f(\vec{a})}{s}$ and $\dd{\Delta}{a_i}\in
\can{\Delta,s_i}$ for every $i\in [1..n]$.
\end{theorem}

\comment{
\begin{proof}
The proof uses three levels of induction: on the function symbols
ordered by $\gtF$, on the sequence of terms to which $f$ is applied,
and on the righthand side structure of the rules defining $f$.
We prove that $\dd{\Delta}{f(\vec{a})}\in \can{\Delta,s}$
by induction on $(\rcint{\Delta}{}(\vec{a}), \vec{a})$ with
$\lexx{(\ge_{stat_f}^S, \lexx{(\arrowstar)})}$ as well-founded order.
\end{proof}
}

\begin{proof}
The proof uses three levels of induction: on the function symbols
ordered by $\gtF$, on the sequence of terms to which $f$ is applied,
and on the righthand side structure of the rules defining $f$.
By induction hypothesis~(1), any $g$ occurring in the rules defining
$f$ satisfies the lemma.

We proceed to prove that $\dd{\Delta}{f(\vec{a})}\in \can{\Delta,s}$
by induction~(2) on $(\rcint{\Delta}{}(\vec{a}), \vec{a})$ with
$\lexx{(\ge_{stat_f}^S, \lexx{(\arrowstar)})}$ as well-founded order.
Since $b=f(\vec{a})$ is a neutral term, by definition of reducibility
candidates, it suffices to prove that every reduct $b'$ of $b$ belongs
to $\can{\Delta,s}$.

If $b$ is not reduced at its root then one $a_i$ is reduced.  Thus,
$b'=f(\vec{a'})$ such that $\vec{a}\arrow \vec{a'}$. As reducibility
candidates are stable by reduction, $\dd{\Delta}{a'_i}\in
\can{\Delta,s_i}$, hence the induction hypothesis~(2) applies since
the interpretation is compatible with reductions.

If $b$ is reduced at its root then $\vec{a}= \vec{c}\theta$ and
$b'=e\theta$ for some terms $\vec{c},e$ and substitution $\theta$ such
that $f(\vec{c})\arrow e$ is the applied rule. $\theta$ is a
well-typed substitution from $\gamR$ to $\Delta$, and $\xi$ is
compatible with $\theta$ since $\dom(\gamR)\cap \varbox= \emptyset$.
We now show that $\theta$ is compatible with $\xi$. Let $x$ be a free
variable of $e$ of type $t$. By definition of the General Schema, $x$
is an accessible subterm of $\vec{c}$. Hence, by
Lemma~\ref{lem-comp-acc-red}, $\dd{\Delta}{x\theta}\in \can{\Delta,t}$
since, for every $i\in [1..n]$, $\dd{\Delta}{c_i\theta}\in
\can{\Delta,s_i}$.

Given an algebraic environment $\Gamma$ containing $\gamR$, let us
show by induction~(3) on the structure of $e\in
\CS{f,_\Gamma}{\vec{c}}$ that, for any well-typed substitution $\theta$
from $\Gamma$ to $\Delta$ compatible with $\xi$, $e\theta\in
\can{\Delta,t}$, provided that $c_i\theta\in \can{\Delta,s_i}$ for
every $i\in [1..n]$.

Base case: either $e$ is accessible in $c_i$, or $e$ is a variable of
$\dom(\Gamma)\setminus \fv(\vec{c})$. In the first case, this results
from Lemma~\ref{lem-comp-acc-red}, and in the second case, this
results from the fact that $\theta$ is compatible with $\xi$. Now, let
us go through the different closure operations of the definition of
$\CS{f,_\Gamma}$.

\begin{list}{}{\parsep=0ex\topsep=0ex\itemsep=0ex
  \leftmargin=5mm\itemindent=-5mm}

\item (i) construction: $e=C(e_1,\ldots,e_p)$ and $\tau(C)=t_1\ra
  \ldots\ra t_p\ra \asort{t}$. $e\theta\in \can{\Delta,\asort{t}}$
  since, by induction hypothesis (3), $e_i\theta\in \can{\Delta,t_i}$.
  
\item (ii) defined application: $e=g(e_1,\ldots,e_p)$ with $\tau(g)=
  t_1\ra \ldots\ra t_p\ra t$ and $g\ltF f$. By induction
  hypothesis~(3), $e_i\theta\in \can{\Delta,t_i}$. Hence, $e\theta\in
  \can{\Delta,t}$, by~\cite{jouannaud97tcs} for first-order function
  symbols, or by induction hypothesis~(1) for higher-order ones, since
  $g\ltF f$.
  
\item (iii) application: $e=\app{u}{v}$. $e\theta\in \can{\Delta,t}$
  since, by induction hypothesis~(3), $u\theta\in \can{\Delta,t'\ra
    t}$ and $v\theta\in \can{\Delta,t'}$.
  
\item (iv) abstraction: $e=\abs{x}{t_1}{u}$ and $t=t_1\tarrow t_2$
  such that $\dedb{\Gamma}{x}{t_1}{u}{t_2}$.  Let $v\in
  \can{\Delta,t_1}$. By induction hypothesis~(3),
  $u\theta\sub{x}{v}\in \can{\Delta,x:t_1,t_2}$. Hence,
  $(\abs{x}{t_1}{u\theta})v\in \can{\Delta,x:t_1,t_2}$ and $e\theta\in
  \can{\Delta,t}$.
  
\item (v) reduction: $e$ is a reduct of a term $u\in
  \CS{f,_\Gamma}{\vec{c}}$.  Since $\ded{\Gamma}{u}{t}$, by induction
  hypothesis~(3), $u\theta\in \can{\Delta,t}$. Since reducibility
  candidates are stable by reduction, $e\theta\in \can{\Delta,t}$.
  
\item (vi) admissible recursive call: $e= f(\vec{c'})$ and
  $\phi_{f,_\Gamma}(\vec{c})= \vec{c} >_{f,_\Gamma}
  \phi_{f,_\Gamma}(\vec{c'})$. The induction hypothesis~(1) applies
  since the interpretation is stable.\qed\\
\end{list}
\end{proof}

\vspace{-2mm}
This achieves the proof of the strong normalization property.
\vspace{-2mm}


%% file: conclusion.tex

\section{Conclusion and future work}
\label{conclusion}

We have defined an extension of the Calculus of Constructions by
higher-order rewrite rules defining uncurried function symbols via the
so called {\em General Schema}~\cite{blanqui98tcssub}, which will
allow a smooth integration in proof assistants like Coq, of function
definitions by pattern-matching on the one hand, and decision
procedures on the other hand. This result extends previous work by
Barbanera et al.~\cite{barbanera97jfp}, by allowing for non-dependent
and non-polymorphic inductive types. In our strong normalization proof
based on Girard's reducibility candidates, we have indeed used a
powerful generalization of the General Schema, of which the recursors
for strictly positive inductive types are an instance, which is an
important step of its own.

Several problems need to be solved to achieve our program, that is to
extend the Coq proof assistant~\cite{coq98} with rewriting facilities.
Firstly, to generalize our results to arbitrary positive inductive
types, for which the type being defined may occur at any positive
position of the argument types of its constructors. Secondly, to
extend the results to dependent and polymorphic inductive types as
defined by Coquand and Paulin in~\cite{coquand88colog}. This is indeed
the same problem, of defining and proving a generalization of the
schema. Thirdly, to allow rewriting at the type level, enabling one to
define types by induction. The corresponding recursor rules are called
strong elimination~\cite{werner94these}. We have already preliminary
results in the latter two directions. Lastly, to accommodate the
$\eta$-rule.  By following~\cite{dicosmo96tcs}, we plan to try the use
of the $\eta$-rule as an expansion, instead of as a reduction. In this
context, it would also be interesting to see to which extent the works
by Nipkow~\cite{nipkow91lics} and Klop~\cite{klop93tcs} on
higher-order rewriting systems could be integrated in our framework.
Fourthly, following~\cite{courant97tlca}, we also want to introduce
modules in our calculus to be able to develop libraries of reusable
parameterized proofs.


%% file: main.bbl
\begin{thebibliography}{10}
\small
\parskip=-1ex

\bibitem{barbanera97jfp}
F.~Barbanera, M.~Fern\'andez, and H.~Geuvers.
\newblock Modularity of strong normalization in the algebraic-$\lambda$-cube.
\newblock {\em Journal of Functional Programming}, 7(6), 1997.

\bibitem{barendregt92jfp}
H.~Barendregt.
\newblock Introduction to generalized type systems.
\newblock {\em Journal of Functional Programming}, 1992.

\bibitem{coq98}
{\em The {C}oq Proof Assistant Reference Manual Version 6.2}.
\newblock INRIA-Rocquencourt-CNRS-Universit\'e Paris Sud-ENS Lyon, 1998.

\bibitem{blanqui98tcssub}
F.~Blanqui, J.-P. Jouannaud, and M.~Okada.
\newblock Inductive {D}ata {T}ype {S}ystems, 1998.

\bibitem{breazu88lics}
V.~Breazu-Tannen.
\newblock Combining algebra and higher-order types.
\newblock In {\em Third IEEE Annual Symposium on Logic in Computer
  Science}, pages 82--90. 1988.

\bibitem{breazu91tcs}
V.~Breazu-Tannen and J.~Gallier.
\newblock Polymorphic rewriting conserves algebraic strong normalization.
\newblock {\em Theoretical Computer Science}, 83(1):3--28, June 1991.

\bibitem{coquand92types}
T.~Coquand.
\newblock Pattern matching with dependent types.
\newblock In B.~Nordstr\"om, K.~Pettersson, G.~Plotkin, editors, {\em
  Workshop on Types for Proofs and Programs}, 1992.

\bibitem{coquand90lf}
T.~Coquand and J.~Gallier.
\newblock A proof of strong normalization for the {T}heory of {C}onstructions
  using a {K}ripke-like interpretation.
\newblock {\em 1st Intl. Workshop on Logical Frameworks}. 1990.

\bibitem{coquand88ic}
T.~Coquand and G.~Huet.
\newblock The {C}alculus of {C}onstructions.
\newblock {\em Information and Computation}, 76:96--120, 1988.

\bibitem{coquand88colog}
T.~Coquand and C.~Paulin-Mohring.
\newblock Inductively defined types.
\newblock In P.~Martin-L\"of and G.~Mints, editors, {\em Proceedings of
  Colog'88}, LNCS 417. Springer-Verlag, 1990.

\bibitem{cornes97these}
C.~Cornes.
\newblock {\em Conception d'un langage de haut niveau de representation de
  preuves: R\'ecurrence par filtrage de motifs; Unification en pr\'esence de
  types inductifs primitifs; Synth\`ese de lemmes d'inversion}.
\newblock PhD thesis, Universit\'e de Paris 7, 1997.

\bibitem{dicosmo96tcs}
R.~Di Cosmo and D.~Kesner.
\newblock Combining algebraic rewriting, extensional lambda calculi, and
  fixpoints.
\newblock {\em Theoretical Computer Science}, 169(2):201--220, 1996.

\bibitem{courant97tlca}
J.~Courant.
\newblock A module calculus for {P}ure {T}ype {S}ystems. {\em TLCA'97}.

\bibitem{dowek98inria3400}
G.~Dowek, T.~Hardin, and C.~Kirchner.
\newblock Theorem proving modulo.
\newblock Technical Report 3400, INRIA, 1998.

\bibitem{gallier90book}
J.~Gallier.
\newblock On {G}irard's ``{C}andidats de {R}\'eductibilit\'e''.
\newblock In P.-G. Odifreddi, editor, {\em Logic and Computer Science}. North
  Holland, 1990.

\bibitem{geuvers94types}
H.~Geuvers.
\newblock A short and flexible proof of strong normalization for the {C}alculus
  of {C}onstructions.
\newblock In P.~Dybjer, B.~Nordstr\"om, and J.~Smith, editors, {\em Selected
  Papers 2nd Intl. Workshop on Types for Proofs and Programs, TYPES'94,
  B{\aa}stad, Sweden, 6--10 June 1994}, volume 996 of {\em LNCS},
  pages 14--38. 1995.

\bibitem{girard88book}
J.-Y. Girard, Y.~Lafont, and P.~Taylor.
\newblock {\em Proofs and Types}.
\newblock Cambridge Tracts in Theoretical Computer Science. Cambridge
  University Press, 1988.

\bibitem{jouannaud97tcs}
J.-P. Jouannaud and M.~Okada.
\newblock {A}bstract {D}ata {T}ype {S}ystems.
\newblock {\em Theoretical Computer Science}, 173(2):349--391, February 1997.

\bibitem{klop93tcs}
J.~W. Klop, V.~van Oostrom, and F.~van Raamsdonk.
\newblock Combinatory reduction systems: introduction and survey.
\newblock {\em Theoretical Computer Science}, 121(1-2):279--308, December 1993.

\bibitem{nipkow91lics}
T.~Nipkow.
\newblock Higher-order critical pairs.
\newblock In {\em Proc. 6th IEEE Symp. Logic in Computer Science, Amsterdam},
  pages 342--349, 1991.

\bibitem{okada89sac}
M.~Okada.
\newblock Strong normalizability for the combined system of the typed lambda
  calculus and an arbitrary convergent term rewrite system.
\newblock In G.~H. Gonnet, editor, {\em Proceedings of the ACM-SIGSAM 1989
  International Symposium on Symbolic and Algebraic Computation}, pages
  357--363. ACM Press, July 1989.

\bibitem{werner94these}
B.~Werner.
\newblock {\em Une Th\'eorie des {C}onstructions {I}nductives}.
\newblock Th\`ese, Universit\'e Paris 7, 1994.

\end{thebibliography}
